\documentclass[prl,twocolumn,a4paper,twoside,showpacs,amsmath,amsfonts]{revtex4}

\usepackage[utf8]{inputenc}
\usepackage[OT1]{fontenc}
\usepackage{graphicx}
\usepackage{stmaryrd} 

\usepackage{pxfonts} 
\usepackage{hyperref}

\begin{document}
\newcommand{\preprintonly}[2][]{#2}	

\newcommand{\arxiv}[1]{arXiv:\href{http://arXiv.org/abs/#1}{#1}}
\newcommand{\aarxiv}[1]{\preprintonly{, \arxiv{#1}}}
\newcommand{\doilink}[1]{http://dx.doi.org/#1}
\newcommand{\JRef}[4][]{\emph{#2} #1 \textbf{#3} #4}
\newcommand{\RMP}[2]{\href{\doilink{10.1103/RevModPhys.#1.#2}}{\JRef{Rev. Mod. Phys.}{#1}{#2}}}
\newcommand{\RMPacc}{\href{http://rmp.aps.org/accepted}{\emph{Rev. Mod. Phys.}}}
\newcommand{\NJP}[3][]{\href{\doilink{10.1088/1367-2630/#2/#1/#3}}{\JRef{New J. Phys.}{#2}{#3}}}
\newcommand{\PRL}[2] {\href{\doilink{10.1103/PhysRevLett.#1.#2}}{\JRef{Phys. Rev. Lett.}{#1}{#2}}}
\newcommand{\PRA}[2] {\href{\doilink{10.1103/PhysRevA.#1.#2}}{\JRef{Phys. Rev. A}{#1}{#2}}}
\newcommand{\QIC}[2]{\href{http://www.rintonpress.com/journals/qiconline.html\#v#1n1}{\JRef{Quant. Inf. Comp.}{#1}{#2}}}
\newcommand{\JCrypt}[2]{\JRef{J. Cryptology}{#1}{#2}}
\newcommand{\Nature}[2]{\JRef[(London)]{Nature}{#1}{#2}}
\newcommand{\PLA}[2]{\JRef{Phys. Lett. A}{#1}{#2}}

\newcommand\ket[1]{\left|#1\right\rangle}
\newcommand\bra[1]{\left\langle#1\right|}
\newcommand\ketbra[1]{\ket{#1}\bra{#1}}
\newcommand\braket[2]{\left\langle #1 | #2 \right\rangle}
\newcommand\isbydef{{:=}}
\newcommand\abs[1]{\left|#1\right|}

\newcommand\Id{{\mathbb I}}
\newcommand{\Proba}[1]{{\mathcal P(#1)}}
\newcommand{\Hamming}[1]{\left\lVert #1 \right\rVert}
\newcommand{\M}{\mathbb M}

\newcommand{\ie}{\emph{i.e.} }
\newcommand{\eg}{\emph{e.g.} }
\newcommand{\etal}{\emph{et al.}}
\newcommand{\prletal}[1]{\preprintonly[ \etal]{#1}}

\title{Robust and Efficient Sifting-Less Quantum Key Distribution Protocols}
\author{Frédéric Grosshans}
\affiliation{Laboratoire de Photonique Quantique et Moléculaire, ENS de Cachan, UMR CNRS 8735, 94235 Cachan cedex, France}
\email{frederic.grosshans@ens-cachan.fr}

\begin{abstract}
	We show that replacing the usual sifting step of the standard 
	quantum-key-distribution protocol BB84 \cite{BB84} by 
	a one-way reverse reconciliation procedure increases its robustness 
	against photon-number-splitting (\textsc{pns}) attacks to the level of the SARG04 
	protocol \cite{SARG1,SARG2} while keeping the raw key-rate of BB84.
	This protocol, which uses the same state and detection than BB84, is the 
	$m=4$ member of a protocol-family  using $m$ polarization states 
	which we introduce here. We show that the robustness of these protocols
	against \textsc{pns} attacks increases exponentially with $m$, and that the
	effective keyrate of optimized weak coherent pulses decreases with the
	transmission $T$ like $T^{1+\frac1{m-2}}$. 
\end{abstract}

\date{\today, version 1.2}


\maketitle


Over the last 25 years, quantum key distribution (QKD)  
has emerged as the main application of quantum information.
In most experimental realizations \cite{SBPCDLP09},
the legitimate partners --- traditionally named Alice and Bob ---
use the BB84 protocol \cite{BB84} with weak-coherent-pulses (\textsc{wcp}), \ie
Alice sends polarized coherent states  to Bob,
and Bob measures their polarization to obtain the raw-key.
Alice and Bob then post-select a subset of the measurement to obtain the 
sifted-key from which the cryptographic key is extracted. 
If Alice sends perfect single-photons, there is no way for an eavesdropper
--- traditionally named Eve ---
to learn anything about the sifted key without introducing errors. 
But, with \textsc{wcp}s, Alice only approximates single-photon, and she sometimes sends
multiphoton pulses, on which Eve can get all the information through 
photon-number-splitting (\textsc{pns}) attack \cite{Luetkenhaus00}.
SARG04 \cite{SARG1, SARG2} showed that, with the same modulation and detection than BB84, 
one can construct 
a protocol more robust against \textsc{pns}, since Eve only gains partial 
information from 2 photons pulse and needs to wait for the rarer 3 photons pulses 
to gain the full information. However, for the same pulse intensity, SARG04's rate
is the half of BB84 at low losses, because of the lower rate of it sifting. 
As shown in \cite{SARG2} 
SARG04's robustness can be increased by using $m$ polarizations instead of 4,
at the price of a lower sifting rate $\propto m^{-3}$.
This article shows that this price is not
necessary, and that it is possible to have the best of both protocols, \ie BB84's 
rate and SARG04's robustness against photon number-splitting attacks. 

BB84 and SARG04 are sifting based protocols \ie protocols where a part of
the data is ``sifted away'' because Alice's state and Bob's measurement are not 
in the ``same basis''. We will look here at sifting-less protocols, 
\ie protocols where this discussion is absent, and therefore, where the 
"wrong-basis" data are kept in the raw-key.

\paragraph{Protocol description.}

Alice randomly choses one linear polarization and sends the corresponding 
phase-randomized
weak coherent pulse (\textsc{wcp}). 
Let $m\ge3$ the total number of 
possible polarizations. To simplify the analysis, we will suppose that
the polarizations are uniformly distributed along a great circle of 
Poincaré's sphere. 
Let $\ket0$ and $\ket1$ be the state of two orthogonally polarized single
 photons.
If the pulse contains $n$ 
photon, Alice sends the state 
$\ket{x,n,m}\isbydef\ket{x\theta_m}^{\otimes n}$ 
with $\theta_m \isbydef\frac{2\pi}m$, 
$x$ uniformly chosen in $\llbracket0,m-1\rrbracket$,
and 
	$\ket\theta 
	\isbydef\tfrac1{\sqrt2}\left(\ket0 + e^{i\theta}\ket1\right)$.
%
If $m=4$, one has the 4 states used in BB84, SARG04 as well as LG09 
\cite{LeverrierGrangier09}.

Bob measures the polarization of the pulses after a propagation into a channel 
of transmission $T$. The public comparison of a small subset of the measurements
allows Alice and Bob to statistically determine the characteristic of the 
channel, namely $T$ and
its qubit error rate (\textsc{qber}).
In this first analysis, we will suppose this statistical evaluation to be exact, 
neglecting the finite size effects \cite{CaiScarani09}.
We will also limit ourselves to the errorless case, where the \textsc{qber} is 0,
excepted in the conclusion where the influence of errors is briefly studied.

There are several possibilities for Bob's measurement.
We will limit Bob's apparatus to single-photon detector based set-ups,
similar to the one used in the BB84
and SARG04 protocols. This will prevent Alice and Bob to extract all the 
information allowed by the Holevo bound $S(X{:}Y)=T\log 2$, or to use 
continuous-variable detection set-up 
\cite{LeverrierGrangier09}. 

Since Bob's measurement is based on single photon detectors, Alice and Bob
need to postselect-away the event when Bob has received no photon \ie when Bob's 
detectors do not click. This can be done by one-way classical communication from 
Bob to Alice. The kept events constitute a fraction $1-e^{-T\mu}\simeq T\mu$
of the sent pulses if the sent \textsc{wcp} have a mean photon number of $\mu$. 
They constitute the raw key, $X$ for Alice and $Y$ for Bob.

When Bob receives a single photon, 
he makes the \textsc{povm} 
$\left\{\tfrac2m\ketbra{y\theta_m+\pi}\right\}_{y\in\llbracket0,m-1\rrbracket}$.
The $\pi$ dephasing doesn't change anything if $m$ is even, 
but increases the mutual information $S(X{:}Y)$ between Alice and Bob when $m$ is 
odd. In particular, it ensures that, for any state sent by Alice, one outcome ($y=x$) 
of Bob's measurement is impossible. 
One can then easily show $S(Y)=\log m$;
\begin{gather}
	\Proba{y|x, m}=\tfrac1m\left(1-\cos(y-x)\theta_ m\right);
		 	\label{eq:CondProb}\displaybreak[1]\\
	S(Y|X)=\log m
		  -\frac1m\sum_{k=0}^{m-1}
			(1 -\cos k\theta_m)\log(1 -\cos k\theta_m);\displaybreak[1]\\
	S(X{:}Y)=\frac1m\sum_{k=0}^{m-1}
			(1 -\cos k\theta_m)\log(1 -\cos k\theta_m).
\end{gather}
The mutual information between Alice and Bob
$S(X{:}Y|m)$ decreases slightly with $m$, 
from $\log\tfrac32=0.5850$ bits for $m=3$ 
to $\tfrac1{2\pi}\int(1 -\cos k\theta)\log(1 -\cos k\theta)d\theta=0.4427$ bits
in the continuous limit $m\rightarrow\infty$.
For $m=4$, we have $S(X{:}Y|m=4)=\frac12\log2$.

When Bob receives more than one 
photon, several detectors can click. This gives him more information than single 
clicks, so neglecting this case , as done above, is pessimistic. 
This corresponds to Bob randomly chosing between the various detection results.

In a reverse reconciliation (\textsc{rr}) scheme \cite{GrosshansGrangier02-Proc, GVAWBCG03},
Alice and Bob can share a common key of length $S(X{:}Y)$ provided Bob sends to 
Alice $S(Y|X)$ bits of information. For example, when $m=4$, 
Bob needs to send 1.5 bits per pulse. This can be done by revealing his 
measurement basis (1 bit/pulse) and using the syndrome of a good erasure correcting
(see \eg \cite[Chapter 50]{MacKay})
code which will be slightly over $\frac12$ bit long per pulse. 
Indeed, when Bob has revealed his basis measurements, Alice knows which bits of $Y$
she knows (the one with the right basis), and the one she does not know (the other ones),
and this corresponds to an erasure channel of rate $\frac12$. 

\paragraph{Eavesdropping.}
Their use of erasure correcting codes 
instead of interactively throwing some bits away 
is at the heart of
the resistance of this protocol against \textsc{pns} attacks : on 2-photon pulses, Eve 
can keep a copy of the pulse sent by Alice, and, even if she knows the basis of 
Bob's measurement, she ignores  whether Alice sent a state in the right basis 
or not. Therefore, in this case, Eve measurement has at best a $25\%$ error-rate, 
giving her at most $h(\tfrac14)=0.1887$ bits of information 
--- where $h(\cdot)$ is the binary entropy ---
while Alice
still has half a bit. The net key rate of 2-photon pulses is then 0.3113 bits.
In BB84, on the contrary, Alice reveals her basis choice, living her on equal
footing with Eve for 2-photons pulses.

Note that when $m$ is even, the above idea for the reconciliation can be
generalized \ie Bob reveals $\log m -\log2$ bits for the basis 
$y \mod {\tfrac m2}$ and use the 
appropriate error correcting code for the remaining information. We are then in 
a situation where Alice has different known error rates 
$\tfrac12(1-\cos(x-y \mod \tfrac m2)\theta_m)$ for different 
bits while Eve only sees the average error rate.
The following paragraphs will study the above affirmations more formally, in
the asymptotic and error-less regime.

Of course, if Alice sends perfect 
single photon pulses, the lack of errors guarantees a perfect secrecy of the 
$S(X{:}Y)$ key.
However, if Alice uses weak coherent pulses (\textsc{wcp}) some attacks become possible 
without introducing errors, 
namely \emph{intercept resend with unambiguous state discrimination (\textsc{irud})}
and \emph{photon number splitting attacks (\textsc{pns})}, as well as a combination of the two.

In any case, since Alice's pulses are phase randomized,
Eve optimal attack starts by a quantum non-demolition measurement of the photon number 
$n$  of Alice's pulse \cite{Luetkenhaus00}. 
The state sent by Alice is then projected onto
\begin{align}
\ket{x,n,m}&=\ket{x\theta_m}^{\otimes n}
	=2^{-\frac n2}(\ket0+e^{ix\theta_m}\ket1)^{\otimes n}\\
	&=2^{-\frac n2}\sum_{b=0}^{2^n-1}e^{i\Hamming b\theta_m}\ket b, 
\end{align}
where $\ket b$ is the tensorial binary development of $b$ and $\Hamming b$ its Hamming 
weight. Note that all terms with the same Hamming weight $w$ modulo $m$ have the 
same phase prefactor $e^{iw\theta_m}$. 
These $\binom{n}{w[m]}$ vectors are orthogonal. We have defined
\begin{equation}
	\binom{n}{w[m]}\isbydef
		\sum_{d=0}^\infty
		\binom{n}{w+dm},
\end{equation}
where we have used the usual convention for the binomial coefficient
 $\binom{n}{w} =0$ for $w>n$.
Let's define, for each $w\in\llbracket0,m-1\rrbracket$,
\begin{equation}
	\ket{w[m]}_n\isbydef
		\frac1{\sqrt{\binom{n}{w[m]}}}
		\sum_{\substack{b\in\llbracket0,2^n-1\rrbracket\\
			b\equiv w[m]}}
		\ket{b}.
\end{equation}
We can then rewrite the state $\ket{x,n,m}$ as
\begin{equation}
	\ket{x,n,m}=
		2^{-\frac n2}
		\sum_{w=0}^{m-1}e^{iw\theta_m}\textstyle{\sqrt{\binom{n}{w[m]}}}\ket{w[m]}_n.
	\label{eq:ketxnm}
\end{equation}

When Eve measures $n$ photons, she can either block the pulse, perform an \textsc{irud} attack
or a \textsc{pns} attack.

\paragraph{\textsc{irud} attacks.}
If Eve makes an \textsc{irud} attack, her success probability is given in 
\cite{CheflesBarnett98} as
\begin{align}
	\Proba{\Delta|m,n}&=2^{-n}m\min_{w\in\llbracket0,m-1\rrbracket}
		\textstyle{\binom{n}{w[m]}}.
\end{align}
This probability is not null iff $n\ge m-1$, and its value 
increases each time $n$ increases by 2. Its first nonzero value is $2^{-m+1}m$ for
$n\in\{m+1,m+2\}$.
If Eves blocks a fraction $b_n$ of the $n$-photon pulses,
these can be a 
the ones where an unambiguous discrimination has failed. 
She can then resend with no error a fraction $u_n$ 
of the original pulses as big as
\begin{equation}
	u_n=
	\max\left(
		\tfrac{\Proba{\Delta|n,m}}{1-\Proba{\Delta|n,m}}b_n;	
		1-b_n	
	\right).
\end{equation}
In other words, she can intercept and resend 
$\frac{\Proba{\Delta|n,m}}{1-\Proba{\Delta|n,m}}$ pulses for each pulse she 
blocks, without introducing any error.

On remaining $p_n=1-b_n-u_n$ pulses, she can perform a \textsc{pns} attack, keeping $n-1$ 
photons
and transmitting the remaining one unperturbed to Bob.
We have
\begin{equation}
	p_n=
	\min\left(
		\tfrac{1-\Proba{\Delta|n,m}-b_n}{1-\Proba{\Delta|n,m}};
		0	
	\right).
\end{equation}

One can construct a Markov chain $Y\leftrightarrow X \rightarrow \ket{x,n-1,m}$,
and since the latter is the state held by Eve when she performs a \textsc{pns} attack,
$S(Y{:}E|n,\text{\textsc{pns}})< S(Y:X)$. 
The inequality is strict because the last transition is not reversible.
In other words, \textsc{pns} attacks without \textsc{irud}
can never reduce the 
net \textsc{rr}-keyrate $K_n=S(Y:X)-S(Y{:}E|n,\text{\textsc{pns}})$ to $0$, 
contrarily to the BB84 protocol.

The net key rate is $0$ when all transmitted pulses can be explained 
by \textsc{irud} attacks, \ie when $\forall n, p_n=0$. 
Let $T_c$ be the critical transmission below which our protocol ceases to work.
At $T_c$, all the $1-e^{-T_c\mu}\simeq T_c\mu$ transmitted pulses correspond to 
the
$e^{-\mu}\sum_{n=0}^{\infty}\frac{\mu^n}{n!}\Proba{\Delta|n,m}$
successful \textsc{irud} attacks. 
We have then
\begin{equation}
	T_c=-\tfrac1\mu\ln\left[1-e^{-\mu}\sum_{n=0}^{\infty}\tfrac{\mu^n}{n!}\Proba{\Delta|n,m}\right]
	\simeq\tfrac{m}{2\cdot m-1!}\left(\tfrac\mu2\right)^{m-2},
	\label{eqTc}
\end{equation}
where the last approximation holds when $\mu\ll1$.
We essentially have $T_c\propto \mu^{m-2}$, showing the exponentially increasing 
robustness  of the protocol for increasing $m$. 
This dependency is 
the same as SARG04,  
but not as BB84,
where $T_c\simeq\frac\mu2$.

\paragraph{\textsc{pns} attack.}
In order to compute the efficiency of the \textsc{pns}-attack, one needs to compute 
the density matrices associated with $n$-photon pulses.
The density matrix corresponding to the state defined in \eqref{eq:ketxnm}
\begin{align}
		\ketbra{x,n,m}=
			2^{-n}
				\sum_{w,w'=0}^{m-1}e^{i(w-w')x\theta_m}\textstyle{\sqrt{\binom{n}{w[m]}\binom{w'[m]}{n}}}
			\notag\\\times
				\ket{w[m]}\bra{w'[m]}
			\\
			=\sum_{D=1-m}^{m-1}
				e^{iDx\theta_m}\M_{D,m,n},
\end{align}
	where we have defined, 
	for any integer $D\in\llbracket1-m,m-1\rrbracket$, the
	(shifted) $m\times m$ diagonal matrix
	\begin{equation}
	\M_{D,m,n}
		\isbydef
			2^{-n}\sum_{w=\min(0,D)}^{m-1+\min(0,D)}
			\textstyle{\sqrt{\binom{w[m]}n\binom{w+D[m]}n}}
			\ket{w[m]}\bra{w+D[m]}.
	\end{equation}

Let $\rho_{n,m}$ be the average $n$-photon state sent by Alice.
One has then
\begin{equation}
	\rho_{n,m}
		=\sum_{x=0}^{m-1}\tfrac1m\ketbra{x,n,m}
		=\M_{0,m,n}.
\end{equation}

When Bob measures $Y=y$, and
Eve keeps $n$ photons, her state conditioned on Bob's measurement is given by
\begin{align}
	\rho_{y,n,m}&=\sum_{x=0}^{m-1}\tfrac1m(1-\cos(x-y)\theta_m)\ketbra{x,n}\\
		&=\M_{0}
			-\tfrac{e^{-iy\theta_m}}2(\M_{m-1}+\M_{-1})
			-\tfrac{e^{iy\theta_m}}2(\M_{-m+1}+\M_1).
\end{align}	
Note that in the above equations, the indices $m$ and $n$ have been omitted for 
$\M_D$ for the sake of simplification.

The Holevo limit of the information Eve can gather on Bob's measurement through
 a collective \textsc{pns} attack is
\begin{align}
 	S(Y{:}E|n,\text{\textsc{pns}})
 		&\isbydef S(E|n,\text{\textsc{pns}}) - S(E|Y,n,\text{\textsc{pns}})\\
 		&=S(\rho_{n-1,m})-S(\rho_{y,n-1,m}).
 \end{align}
 These entropies are easily computed numerically and
 decrease slowly with $n$.
 
They are independent of $m$ iff
$n\le m-1$, which means that the corresponding $S(Y{:}E)$ will also be identical 
in this case. In other words, the information leaked to Eve in $m$-state protocols 
are identical to the 
continuous $m\rightarrow\infty$ limit for $n$-photon pulses when $n\le m-2$, and
the only difference at $n=m-1$ comes from the \textsc{irud} attack.

\begin{figure}
	\includegraphics[width=\columnwidth]{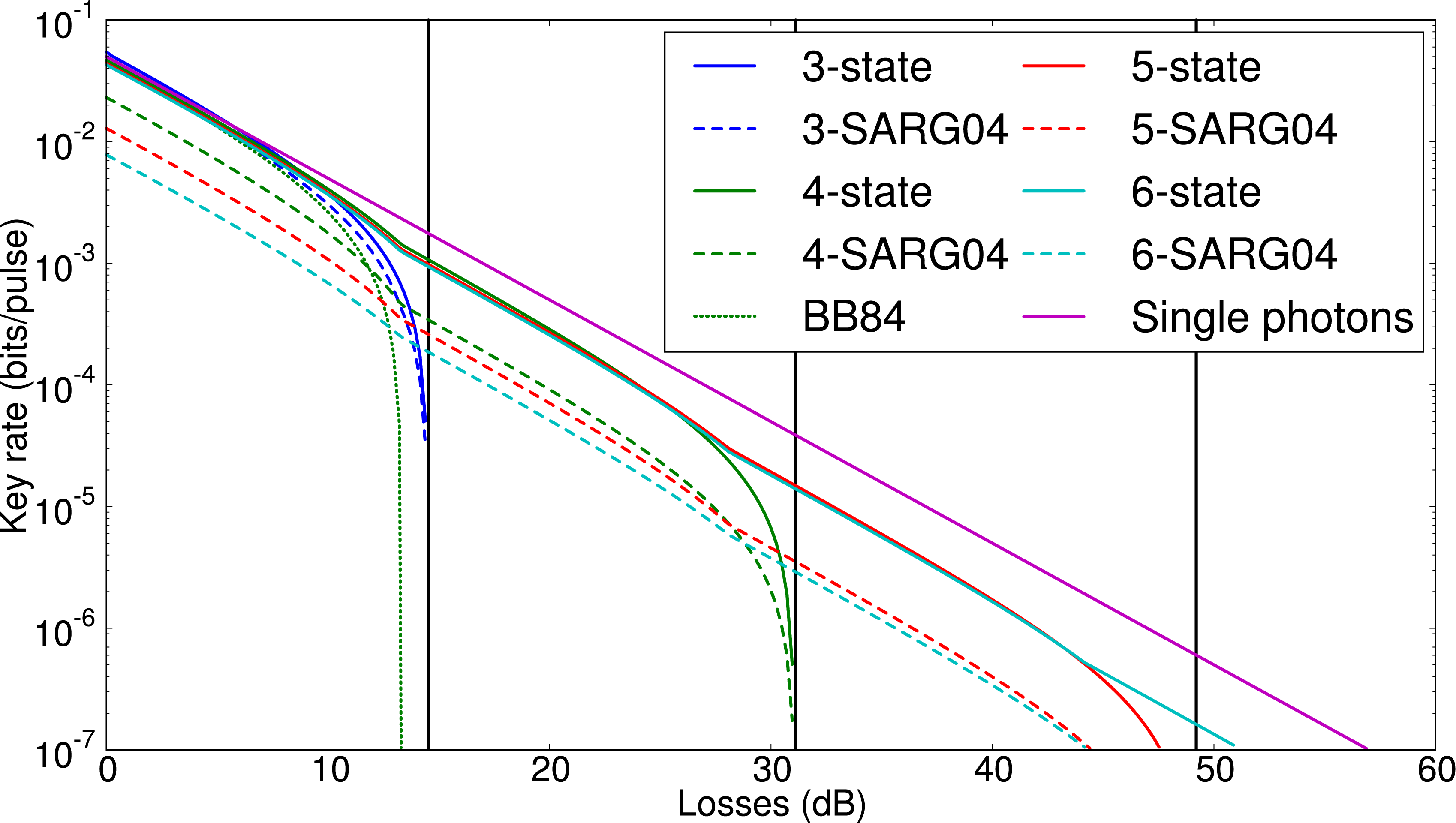}
	\caption{\label{FigKeyRates}
	Key rates of the $m$-states protocols compared to $m$-state SARG04 and BB84
	with \textsc{wcp}s for $\mu=0.1$ and $m\in\llbracket3,6\rrbracket$. 
	The vertical lines represent the values of $T_c$ given by \eqref{eqTc}.}
\end{figure}
\begin{figure}
	\includegraphics[width=\columnwidth]{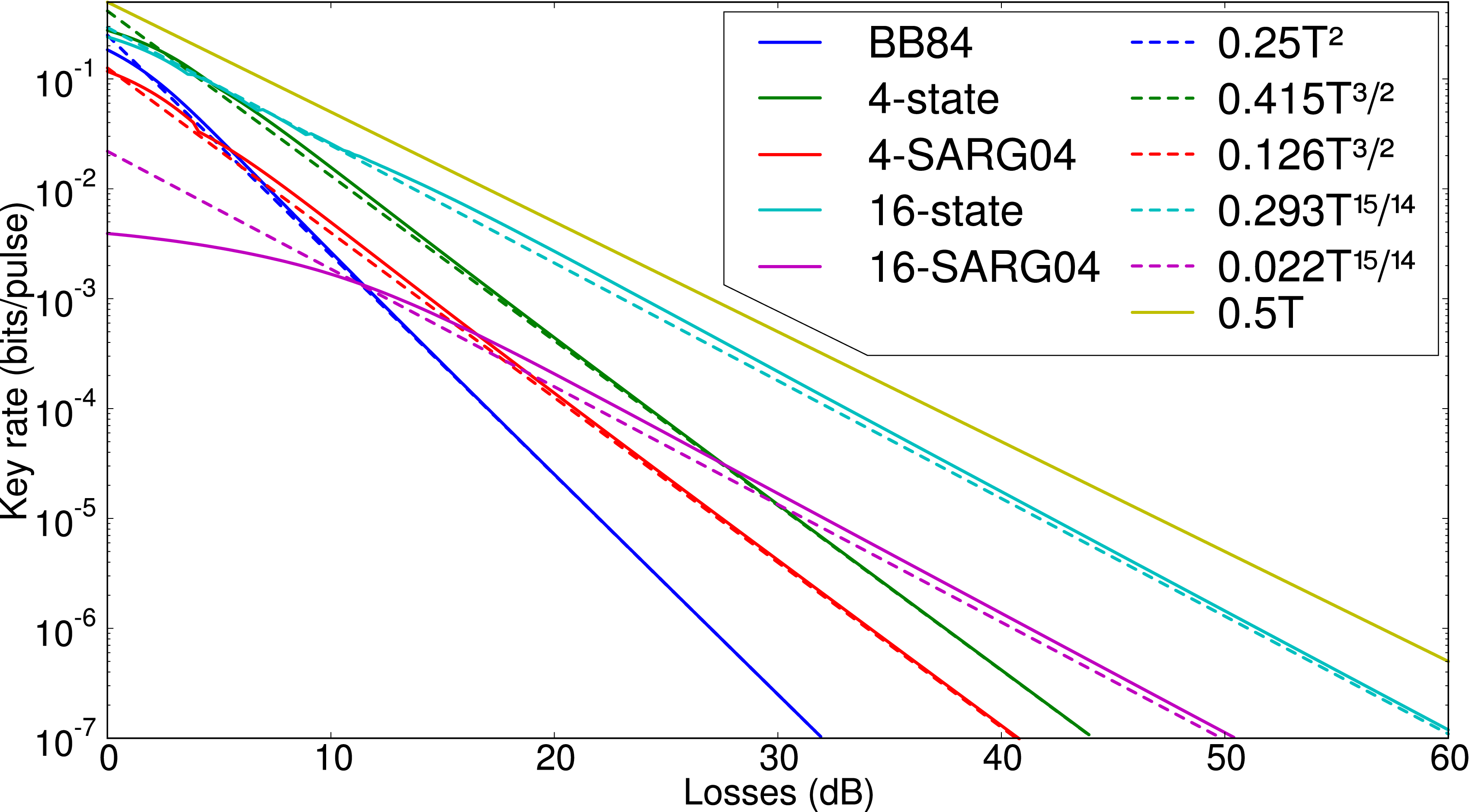}
	\caption{\label{FigOptKeyRates}
	Key rates with optimized $\mu$ for BB84, the $m$-states protocol, 
	$m$-state for $m=4$ and $m=16$. 
	}
\end{figure}
\paragraph{Key-Rate.}
The net key rate $K(T,\mu)$ for
\textsc{wcp} with $\mu$ photons/pulse on average
is therefore
\begin{align}
	K(T,\mu)&=\sum_{n=1}^{\infty}e^{-\mu}\frac{\mu^n}{n!}p_n K_n 
		\text{ with }K_n\isbydef S(X{:}Y)-(Y{:}E|n,\text{\textsc{pns}})\\
	&=\sum_{n=1}^{\infty}e^{-\mu}\frac{\mu^n}{n!}K_n -
		\sum_{n=1}^{\infty}e^{-\mu}\frac{\mu^n}{n!}b_n\frac{K_n}{1-\Proba{\Delta|n,m}}.
\end{align}
When $T\ge T_c$
the optimal attack is for Eve to block the pulses with 
the biggest values of $\frac{K_n}{1-\Proba{\Delta|n,m}}$. This 
corresponds only roughly to the pulses with the lowest photon number. 
The corresponding rates for fixed $\mu=0.1$ are shown in figure \ref{FigKeyRates}.

One can also numerically optimize $\mu$ for each value of the transmission $T$,
as shown in figure \ref{FigOptKeyRates}.
If the optimal key rate is achieved close to $T_c$, we have,
for $\mu\ll1$, 
\begin{align}
	K&\simeq K'_{m-1}\left(T\mu - \Proba{\Delta|m-1,m}\tfrac{\mu^{m-1}}{m-1!}\right)\\
\intertext{with $K'_{m-1}$ being the $(m-1)$th value of the 
$\frac{K_n}{1-\Proba{\Delta,n,m}}$ coefficients in decreasing order. 
Optimizing this quantity for $\mu$ is straightforward and gives
}
	\mu_{\text{opt}}&\simeq2 \left(\tfrac{2\cdot m-2!}{m}\right)^{\frac1{m-2}}  T^{\frac1{m-2}}\\
	K_{\text{opt}}&\simeq
			K'_{m-1}
			\tfrac{2}{m-1} 
			\left(\tfrac{2\cdot m-2!}{m}\right)^{\frac1{m-2}}
			T^{1+\frac1{m-2}}
\end{align}
\ie the key rate essentially varies as  
$K\propto T^{1+\frac1{m-2}}$ with a prefactor which slowly 
decreases with $m$. This approximation seems in agreement
with numerical results, at least for reasonably low $m$ (below 16).
The bigger $m$ is, the closer one is to the ideal single-photon case,
where $K=\frac T2\log 2$.

\paragraph{Conclusion}
The sifting-less protocols described here are as efficient as BB84 and more
robust against \textsc{pns}-attack.
This robustness 
lies in the preservation of non-orthogonality of the sent-states
by the lack of sifting.

Furthermore, this also allows to 
extract a reasonable key for high $m$, while benefiting of the
robustness brought by the increased overlap of the sent states, on the
contrary to the $m$-state SARG04 variant, which while robust, have a sifting 
factor $\propto m^{-3}$ \cite{SARG2}.

The most robust variant limit of this protocol is the limit of 
continuous phase modulation $m\rightarrow\infty$, which actually prevents
the \textsc{irud} attack. It is straightforward to show that replacing the $m$-state \textsc{povm}
used in the above description by the simpler 4-State \textsc{povm} used in standard BB84
does not change the
key-rate in this limit. 

Before using this protocol, we still need to investigate its security 
in presence of a non-zero \textsc{qber}. 
For perfect single photons and a \textsc{qber} $\epsilon$, 
one can bound Eve's information by writing the state
shared by Alice, Bob and Eve under the form \cite{SBPCDLP09}
$\ket{\Psi_{ABE}}
	=\sqrt{\lambda_1}\ket{\Phi^+}\ket{E_1}
	+\sqrt{\lambda_2}\ket{\Phi^-}\ket{E_2}
	+\sqrt{\lambda_3}\ket{\Phi^+}\ket{E_3}
	+\sqrt{\lambda_4}\ket{\Phi^-}\ket{E_4}$,
and optimizing Eve's Holevo information $S(Y{:}E)$. 
%
One then straightforwardly find $S(Y{:}E|n=1,\epsilon)=h(\epsilon)$.
For $m=4$, we have 
$S(X{:}Y)=\tfrac12(\log2-h(\epsilon))$, which gives a net key rate 
$K=\frac12(\log2 - 3 h(\epsilon))$, cancelling for a \textsc{qber}  $\epsilon=6.14\%$.
The expression is less elegant for other values of $m$, but the critical value of
$\epsilon$ does not change much, varying between 6.89\% for $m=3$ and
5.93\% for $m\rightarrow\infty$. 
Of course, for a practical application of these protocols, the combination of 
\textsc{qber} and \textsc{pns} attacks still needs to be investigated, 
as well as finite-size effects 
\cite{CaiScarani09}.

Another direction worth investigating would be an unbalanced version of our 
protocol, similar to BB84 with biased basis choice \cite{LoChauArdehali98-05}, allowing 
to double the key rate to $\sim1$ bit/pulse instead of $\sim.5$ in the low-loss 
regime.

\begin{acknowledgments}
I thank Valerio Scarani, for bringing the problem of the optimal
sifting of the states used in BB84 and SARG04 to my attention during a visit
at the Centre for Quantum Technologies at the National University of Singapore.
This research has been funded the European Union under the \textsc{equind}
(project \textsc{ist-034368}) and \textsc{nedqit} (\textsc{eranet Nano-Sci}) projects, 
and by the French Agence Nationale de la Recherche \textsc{prospiq} project
(project \textsc{anr-06-nano-041}).

\end{acknowledgments}


\begin{thebibliography}{99}

\bibitem{BB84}
	Charles H. Bennett and Gilles Brassard,
	\emph{``Quantum Cryptography: Public Key Distribution and Coin Tossing''},
	in \emph{Proceedings IEEE International Conference on Computers, Systems 
	and Signal Processing, Bangalore, India} 
	(IEEE, New York, 1984), pp. 175--9
\bibitem{SARG1}
	Valerio Scarani, Antonio Acín, Grégoire Ribordy and Nicolas Gisin,
	\preprintonly{``Quantum Cryptography Protocols Robust against 
		Photon Number Splitting Attacks for Weak Laser Pulse Implementations'',}
	\PRL{92}{057901} (2004)%
	\aarxiv{quant-ph/0211131}.
\bibitem{SARG2} 
	Antonio Acín, Nicolas Gisin, and Valerio Scarani,
    \preprintonly{``Coherent-pulse implementations of quantum cryptography 
     	protocols resistant to photon-number-splitting attacks'',}
	\PRA{69}{012309}(2004)%
	\aarxiv{quant-ph/0302037}.
\bibitem{SBPCDLP09}
	Valerio Scarani, Helle Bechmann-Pasquinucci, Nicolas J. Cerf , 
		Miloslav Dušek, Norbert Lütkenhaus, Momtchil Peev,
	\preprintonly{``The Security of Practical Quantum Key Distribution''},
	\arxiv{0802.4155} (2008). 
	To appear in \RMPacc.
\bibitem{Luetkenhaus00}
	Norbert Lütkenhaus,
	\preprintonly{``Security against individual attacks for realistic quantum key distribution'',}
	\PRA{61}{052304} (2000)%
	\aarxiv{quant-ph/9910093}.
\bibitem{LeverrierGrangier09}
	Anthony Leverrier and Philipper Grangier,
	\preprintonly{``Unconditional Security Proof of Long-Distance 
		Continuous-Variable Quantum Key Distribution with Discrete Modulation'',}
	\PRL{102}{180504} (2009)%
	\aarxiv{0812.4246}.
\bibitem{CaiScarani09}
	Raymond Y.Q. Cai, Valerio Scarani,
	\preprint{``Finite key analysis for practical implementations of quantum key 
		distribution'',} 
	\NJP[4]{11}{045024}%
	\aarxiv{0811.2628}.
\bibitem{GrosshansGrangier02-Proc} 
	F. Grosshans and Ph. Grangier, 
	in \emph{\preprintonly[Proc. 6th Int. Conf. on Quant. Comm. Meas. and Comp.]%
		{Proc. 6th International Conference
		on Quantum Communications, Measurement, and Computing}}, 
	edited by J. H. Shapiro and J. O. Hirota   (Rinton
  	Press, December 2002), p. 351%
  	\aarxiv{quant-ph/0204127}.
\bibitem{GVAWBCG03}
	Frédéric Grosshans\prletal{, Gilles Van Assche, Jérôme Wenger, Rosa Brouri,
  		Nicolas J. Cerf and Philippe Grangier}, 
  	\Nature{421}{238} (2003)%
	\aarxiv{quant-ph/0312016}.
\bibitem{MacKay}
	David J. C. MacKay,
	``Information Theory, Inference and Learning Algorithms'', 
	Cambridge University Press (2003).
	\url{http://www.inference.phy.cam.ac.uk/mackay/itila}.
\bibitem{CheflesBarnett98}
	Anthony~Chefles and Stephen~M.~Barnett,
	\preprintonly{``Optimum unambiguous discrimination between linearly independent 
		symmetric states'',}
	\PLA{250}{223}(1998).
\bibitem{LoChauArdehali98-05}
	Hoi-Kwong Lo, H. F. Chau, M. Ardehali,
	\preprintonly{``Efficient Quantum Key Distribution Scheme and a Proof of Its 
		Unconditional Security'',} 
	\JCrypt{18}{133} (2005)%
	\aarxiv{quant-ph/0011056};
	and \preprintonly{``Efficient Quantum Key Distribution Scheme'',} 
	\arxiv{quant-ph/9803007} (1998).
\end{thebibliography}
\end{document}